\documentclass{iau307}
\usepackage{graphicx}
\usepackage{natbib}
\usepackage{url}
\usepackage{dtklogos}
\bibpunct{(}{)}{;}{a}{}{,}

\title[Properties of single WO stars] 
{The properties of single WO stars}

\author[F. Tramper, S. Straal, et al.]   
{F. Tramper$^1$,
 S.M. Straal$^1$,
 G. Gr\"afener$^2$,
 L. Kaper$^1$,
 A. de Koter$^{1,3}$,
 N. Langer$^4$,
 H. Sana$^5$
 \and J.S. Vink$^2$ }

\affiliation{$^1$Anton Pannekoek Institute for Astronomy, University of Amsterdam, PO Box 94249, 1090 GE Amsterdam, The Netherlands \\ email: {\tt F.Tramper@uva.nl} \\[\affilskip]
$^2$Armagh Observatory, College Hill, BT61 9DG Armagh, Northern Ireland, UK \\[\affilskip]
$^3$Instituut voor Sterrenkunde, KU Leuven, Celestijnenlaan 200D, 3001 Leuven, Belgium \\[\affilskip]
$^4$ Argelander Institut f\"ur Astronomie, University of Bonn, Auf dem H\"ugel 71, D-53121 Bonn, Germany\\[\affilskip]
$^5$ESA / Space Telescope Science Institute, 3700 San Martin Drive, Baltimore, MD 21218, USA}

\pubyear{2014}
\volume{307} 
\pagerange{}
\jname{New windows on massive stars: asteroseismology, interferometry, and spectropolarimetry}
\editors{G. Meynet, C. Georgy, J.H. Groh \& Ph. Stee, eds.}

\begin{document}

\maketitle

\begin{abstract}
The enigmatic oxygen sequence Wolf-Rayet (WO) stars represent a very late stage in massive star evolution, although their exact nature is still under debate. The spectra of most of the WO stars have never been analysed through detailed modelling with a non-local thermodynamic equilibrium expanding atmosphere code. Here we present preliminary results of the first homogeneous analysis of the (apparently) single WOs. 
\end{abstract}

\firstsection 
\section{Introduction}
The nature of the very rare WO stars is still under debate. Although it is clear that they are highly evolved massive stars, their evolutionary connection to the more common carbon sequence Wolf-Rayet (WC) stars remains unclear. To unravel their nature, we are undertaking a project targeting all the known WO stars. Here we present preliminary results of the first homogeneous analysis of the single WOs.

\section{Observational sample and spectral classification}

\begin{figure}[b]
\begin{center}
\includegraphics[width=0.95\textwidth]{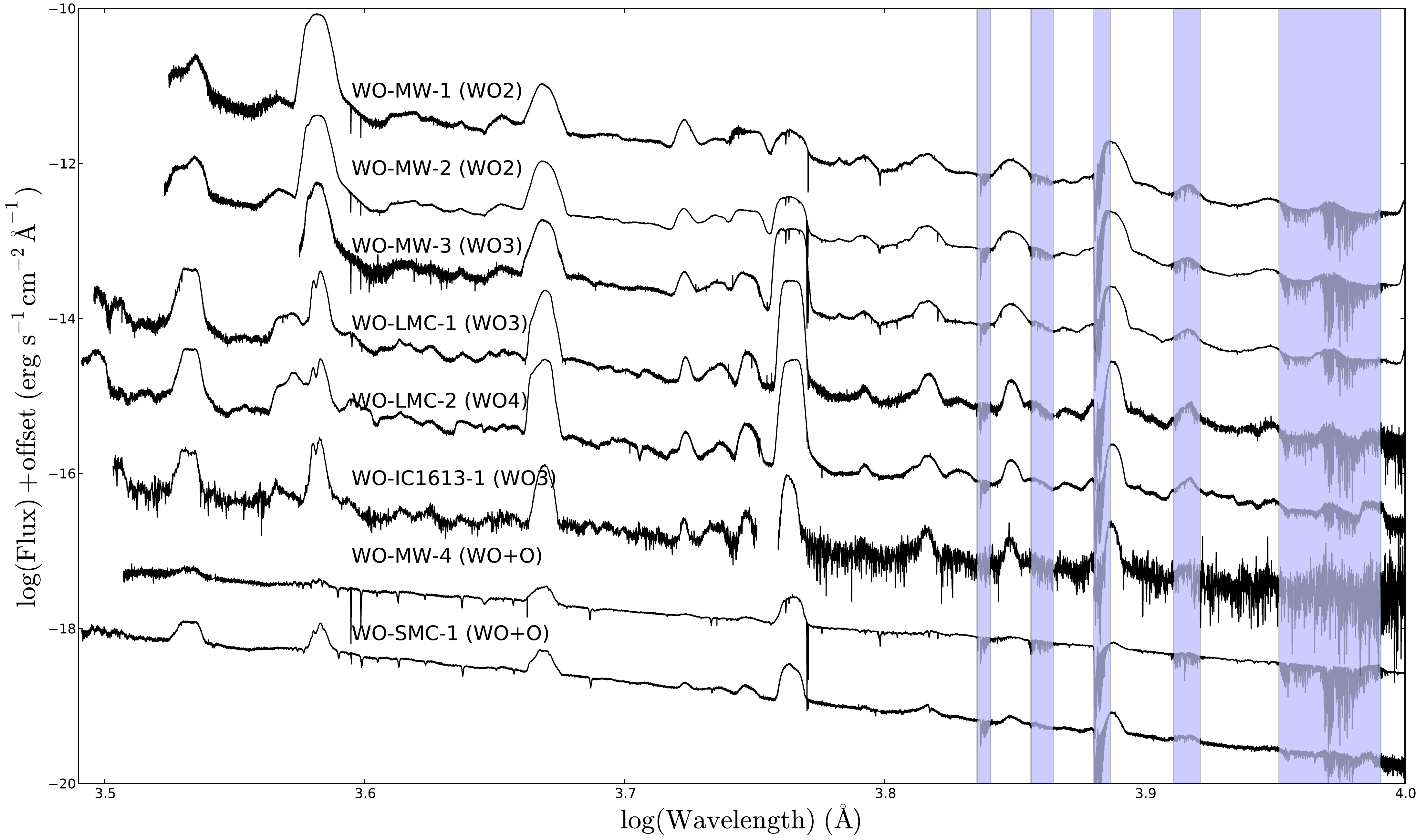} 
\caption{Dereddened, flux-calibrated spectra of the WO stars in the 3\,000-10\,000 \AA \ wavelength range. The shaded areas indicate regions with telluric features. Labels are defined in Table~\ref{tab:results}.}
\label{fig:atlas}
\end{center}
\end{figure}

We have obtained the near-ultraviolet to near-infrared spectra of all known WO stars using the X-Shooter spectrograph on ESO's Very Large Telescope \citep[except for the recently discovered WO star in the LMC;][]{massey2014}. The spectra cover a wavelength range from 3\,000 to 25\,000 \AA \ at a resolving power of $R \sim 8000$. The sample consists of three stars in the Milky Way (MW), two in the Large Magellanic Cloud (LMC), and one in IC~1613. The two WO stars in a binary system have also been observed, but are not analysed in this work. Figure~\ref{fig:atlas} shows the X-Shooter spectra of all WO stars.

\begin{table}[t]
\begin{center}
\caption{Spectral types and preliminary parameters of the single WO stars.}
\label{tab:results}
\begin{tabular}{l l c c c c c}\hline
\textbf{ID} & \textbf{ID} \scriptsize{\textit{(figure})} & \textbf{SpT} & \textbf{log\,\textit{L}} & \textbf{\textit{T}}$_{*}$ & \textbf{X}$_{\mathbf{C}}$ & \textbf{X}$_{\mathbf{O}}$\\
& 	& & $(L_{\odot})$	& (kK)	\\
\hline
WR102	&	WO-MW-1	&	WO2	& 	5.45	&	210	&	0.62		&	0.25 \\
WR142	&	WO-MW-2	&	WO2	&	5.63	&	200	&	0.54		&	0.21 \\
WR93b	&	WO-MW-3	&	WO3	&	5.30	&	160	&	0.53		&	0.18 \\	
BAT99-123	& WO-LMC-1	& WO3& 	5.26 &	170	& 	0.55		& 	0.15 \\
LH41-1042	& WO-LMC-2	& WO4& 	5.26 & 	150	&	0.60		&	0.18 \\
DR1		&	WO-IC1613-1	& WO3& 	5.68	& 	150	&	0.46		&	0.10 \\
\hline
\end{tabular}
\end{center}
\scriptsize{
{\it Notes:} Results for DR1 from \cite{tramper2013}. Values for all other stars are preliminary and subject to change.}\\
\end{table}

The spectral type of the single stars is determined using the classification criteria from \cite{crowther1998}. These are based on the equivalent width ratios of O{\sc vi} 3811-34 \AA \ / O{\sc v} 5590 \AA \ and O{\sc vi} 3811-24 \AA / C{\sc iv} 5801-12 \AA, as well as the full width at half maximum of C{\sc iv} 5801-12 \AA. The derived spectral types are given in Table~\ref{tab:results}.

\section{Modelling and conclusions\label{sec:modeling}}
We model the X-Shooter spectra using {\sc cmfgen} \citep{hillier1998} following the method described in \cite{tramper2013}. While the observed strong O{\sc vi} 3811-34 \AA \ emission is not (fully) reproduced, the rest of the spectrum is very well represented by our models. Preliminary parameters derived from the modelling are given in Table~\ref{tab:results}. Note that the models are currently still being refined, and these values are subject to small changes.

The surface abundances of the WO stars indicate that they have at least burned two-third of the helium in their core, and are expected to explode as type Ic supernovae within $10^5$ years. Most evolved is WR102, which may have already exhausted the helium in its core and will likely end its life in less than $10^4$ years. Compared to WC stars, the WO stars are hotter and show higher surface abundances of carbon and oxygen. 

\bibliographystyle{iau307}
\bibliography{MyBiblio_Tramper}

\begin{thebibliography}{}

\bibitem[\protect\astroncite{{Crowther} et~al.}{1998}]{crowther1998}
{Crowther}, P.~A., {De Marco}, O., \& {Barlow}, M.~J. 1998,
\newblock {\em \mnras} 296, 367

\bibitem[\protect\astroncite{{Hillier} \& {Miller}}{1998}]{hillier1998}
{Hillier}, D.~J. \& {Miller}, D.~L. 1998,
\newblock {\em \apj} 496, 407

\bibitem[\protect\astroncite{{Massey} et~al.}{2014}]{massey2014}
{Massey}, P., {Neugent}, K.~F., {Morrell}, N., \& {Hillier}, D.~J. 2014,
\newblock {\em \apj} 788, 83

\bibitem[\protect\astroncite{{Tramper} et~al.}{2013}]{tramper2013}
{Tramper}, F., {Gr{\"a}fener}, G., {Hartoog}, O.~E., {et~al.} 2013,
\newblock {\em \aap} 559, A72

\end{thebibliography}

\end{document}